\documentclass[twocolumn,prl,aps,longbibliography]{revtex4-2}
\usepackage{xr-hyper}  
\usepackage{hyperref} 
\usepackage{graphicx}
\usepackage{rotating}
\usepackage{array}
\usepackage{xcolor}
\usepackage{amsmath}
\usepackage{bm}
\usepackage{multirow}
\usepackage{setspace}
\usepackage{braket}
\usepackage{epstopdf}
\usepackage{siunitx}
\usepackage{chemformula}
\usepackage{gensymb}
\usepackage[version=3]{mhchem} 
\usepackage{makecell}
\usepackage{soul}
\usepackage{overpic}

\usepackage{pdfpages}
\usepackage{pgffor}

\makeatletter
\AtBeginDocument{\let\LS@rot\@undefined}
\makeatother

\DeclareUnicodeCharacter{03C0}{\ensuremath{\pi}}

\begin{document}

\title{Generalizing Density Functional Theory with a Particle-Hole Expansion}

\author{Shurui Zhao, Madhubani Mukherjee, and David A. Mazziotti}

\email{damazz@uchicago.edu}

\affiliation{Department of Chemistry and The James Franck Institute, The University of Chicago, Chicago, Illinois 60637, USA}

\date{Submitted October 29, 2026}

\begin{abstract}
\textcolor{black}{Strong electron correlation remains difficult to describe within conventional density functional approximations, particularly when natural-orbital occupations become strongly fractional. We introduce a systematic particle-hole expansion that generalizes density functional theory (DFT) through an explicit functional of the one-electron reduced density matrix. The previously derived cumulant-based correction emerges as the leading term and the thermal (entropic) correction as an infinite-order member whose coefficients are fixed by a single parameter. The expansion can be formulated as a moment problem solved by semidefinite programming with computational scaling comparable to DFT.  We show that both the first- and second-order corrections yield closed-form natural-orbital occupations.  Applications to periacenes and Fe-porphyrin demonstrate that the second-order correction substantially reduces excessive orbital fractionalization, yielding natural-orbital occupations and fundamental gaps in improved agreement with correlated calculations.}
\end{abstract}

\maketitle

\noindent {\em Introduction}---The accurate and efficient treatment of strong electron correlation remains one of the central challenges in electronic structure theory. Although density functional theory (DFT)~\cite{KS_DFT,Hohenberg_Kohn_1964,Parr_Yang_DFT_1994,Dreizler_Gross_DFT_1990,Jones_DFT_2015} has been remarkably successful for many-electron systems, conventional density functional approximations often struggle with the static correlation~\cite{Huang_Kais_2005, BenavidesRiveros_2017} associated with near-degenerate electronic states and
the associated fractional orbital occupations~\cite{WYang_SC_2008,Hollett_Gill_2011,Chai_DFT_Fractional_2012,Fuchs_StaticCorrelation_2005}. Such strongly correlated electronic structures play an important role in transition-metal chemistry~\cite{StrongCorr_metal}, high-temperature superconductivity~\cite{SuperC_1,Sharma_RDMFT_Mott_2013}, and quantum spin liquids~\cite{qsl_prr_2025}. Accurate descriptions of these systems generally require many-body or multireference electronic structure methods, whose computational cost can limit applications to large molecules and extended systems.

In this Letter, we present a reduced-density-matrix functional theory (RDMFT)---particle-hole RDMFT (ph-RDMFT)---that systematically generalizes existing density functional approximations through a particle-hole expansion of the 1-electron reduced density matrix (1-RDM).  The first term of the expansion agrees with the recent correction derived from the two-electron reduced density matrix (2-RDM) and its cumulant~\cite{Gibney_PRL_2023, Gibney_PRA_2024, Gibney_JPCL_2022, Gibney_JCTC_2022, gibney_jpcl_2021}.  The remainder of the expansion provides a complete ansatz for the additive, particle-hole-invariant dependence of the functional on the natural-orbital occupations, with systematically increasing flexibility for strongly correlated systems.  We show that the expansion can be expressed as a moment expansion of the density matrix that is solvable by semidefinite programming~\cite{Mazziotti_SDP_2}. \textcolor{black}{For the convex second-order correction, the semidefinite program reduces to a closed-form expression for the natural-orbital occupations.  Thermal (entropic) approaches~\cite{Mermin_1965, Chai_DFT_Fractional_2012,JWang_aps_2022, Gibney_JCTC_2022, Ramos_Pavanello_2019, Martinez_Pavanello_2023, Matz_thermal_2026, Han_FTACRDMFT_2026, Graf_temperature_2026}, which add an entropy of the natural-orbital occupations to a mean-field or DFA energy, fall within the particle-hole expansion as an infinite-order case whose coefficients are fixed by a single temperature-like parameter.}  Moreover, we implement the expansion through second order in the particle-hole product and establish bounds on the first- and second-order coefficients from the convexity of the functional.  We demonstrate the particle-hole expansion for molecular bond dissociation, conjugated $\pi$-electron systems with increasing polyradical character, and transition-metal porphyrins with open-shell metal centers.

\noindent {\em Theory---}The ground-state electronic energy can be written as a functional of the 1-RDM~\cite{HK_RDMFT_org_1975,Levy_RDM_1979,Valone_1980,Muller_1984,Goedecker_Umrigar_1998,Schilling_PRL_2019,Mazziotti_Geminal_2000}, ${}^{1}D$, as
\begin{equation}
E[{}^{1}D]
=
\mathrm{Tr}\!\left[{}^{1}H\,{}^{1}D\right]
+
U[{}^{1}D],
\end{equation}
where ${}^{1}H$ contains the kinetic-energy and electron--nuclear attraction operators and $U[{}^{1}D]$ is the universal electron--electron interaction functional. We approximate the universal functional by combining a density functional approximation (DFA) with an explicit 1-RDM-dependent correction~\cite{Gibney_PRL_2023, Gibney_PRA_2024, Gibney_JPCL_2022, Gibney_JCTC_2022, gibney_jpcl_2021},
\begin{equation}
U[{}^{1}D]
=
F_{\mathrm{DFA}}[\rho]
-
T_s[\rho]
+
C[{}^{1}D],
\end{equation}
where $F_{\mathrm{DFA}}[\rho]$ is the universal DFA functional, $T_s[\rho]$ is the noninteracting kinetic-energy functional, and $C[{}^{1}D]$ is the density-matrix correction developed below. The subtraction of $T_s[\rho]$ removes the kinetic-energy contribution already included explicitly through the 1-body Hamiltonian.  Because the resulting functional depends explicitly on the 1-RDM, this construction defines an RDMFT. In contrast to conventional RDMFT, however, the present approach retains the exchange-correlation functional of an existing DFA and augments it with the explicit 1-RDM correction $C[{}^{1}D]$.

The particle and hole 1-RDMs are defined by
\begin{align}
{}^{1}D^{p}_{q}
& =
\langle \Psi|\hat{a}^{\dagger}_{p}\hat{a}^{}_{q}|\Psi\rangle \\
{}^{1}Q^{q}_{p}
& =
\langle \Psi|\hat{a}^{}_{q}\hat{a}^{\dagger}_{p}|\Psi\rangle
=
\delta^{p}_{q}-{}^{1}D^{p}_{q}.
\end{align}
The ensemble $N$-representability conditions~\cite{Coleman1963,Valone_1980,Hammond_Mazziotti_2005,Mazziotti_Nrep_2012,
Mazziotti_Nrep_PRA_2012,Mazziotti_Nrep_2023,Mazziotti.2026,Mazziotti_V2RDM_2016} on the 1-RDM require \textcolor{black}{${}^{1}D\succeq0$ and ${}^{1}Q\succeq0$,} together with the normalization condition $\mathrm{Tr}({}^{1}D)=N$. These conditions constrain the eigenvalues of ${}^{1}D$, the natural-orbital occupation numbers, to lie between zero and one\textcolor{black}{; we write the theory for spin orbitals, while the closed-shell calculations use the equivalent spin-summed 1-RDM with occupations in $[0,2]$ and ${}^{1}Q=2I-{}^{1}D$, which rescales the coefficients.}

In previous work, we derived a 1-RDM correction to density functional
approximations from the two-electron reduced density matrix (2-RDM) and
its cumulant~\cite{Gibney_JPCL_2022,Gibney_PRL_2023}. The resulting
correction is proportional to the particle-hole product
${}^{1}D{}^{1}Q$. More generally, any polynomial occupation-number
correction that vanishes for idempotent 1-RDMs can be expressed in the
form
\begin{equation}
C[{}^{1}D]
=
\sum_{m=0}^{\infty}
a_m\,\mathrm{Tr}\!\left[
{}^{1}D{}^{1}Q
\left({}^{1}D-{}^{1}Q\right)^m
\right].
\label{eq:am}
\end{equation}
The use of a single trace makes each term additive for noninteracting
subsystems and hence size extensive; disconnected products of traces
are therefore not included. Under particle-hole interchange,
${}^{1}D\leftrightarrow{}^{1}Q$, the even powers are invariant whereas
the odd powers change sign. Restricting the correction to be
particle-hole invariant therefore retains only the even powers\textcolor{black}{, and because $({}^{1}D-{}^{1}Q)^2=I-4\,{}^{1}D{}^{1}Q$, any finite truncation can be written in powers of the particle-hole product, $\sum_{p}c_p\,\mathrm{Tr}[({}^{1}D{}^{1}Q)^p]$.}
Thus, the cumulant-derived correction is the leading term of a
systematic particle-hole expansion of the occupation-number dependence
of the functional.

Truncating the particle-hole-invariant expansion at second order in ${}^{1}D{}^{1}Q$, which gives a quartic polynomial in ${}^{1}D$, yields the two-parameter model
\begin{equation}
C[{}^{1}D]
=
-w_0\,\mathrm{Tr}\!\left({}^{1}D{}^{1}Q\right)
+
u_0\,\mathrm{Tr}\!\left[
\left({}^{1}D{}^{1}Q\right)^2
\right],
\end{equation}
where $w_0$ and $u_0$ are the first- and second-order expansion coefficients, respectively. Convexity of this correction places a bound on the two coefficients. For a natural-orbital occupation $n$, the corresponding contribution to the correction is
\begin{equation}
c(n)
=
-w_0 n(1-n)
+
u_0 n^2(1-n)^2 .
\end{equation}
Using the second derivative, we find that the minimum curvature occurs at $n=1/2$, where $c''(1/2)=2w_0-u_0$. Consequently, for nonnegative coefficients satisfying
$0\leq u_0\leq 2w_0$ the second-order particle-hole correction is convex over the full occupation interval $0\leq n\leq1$. \textcolor{black}{Convexity of $c$ on $[0,1]$ also makes the trace functional $C[{}^{1}D]=\mathrm{Tr}\,c({}^{1}D)$ convex in ${}^{1}D$.}

The second-order particle-hole correction can be expressed through moments of the 1-RDM.  The corresponding moment matrix, containing powers of ${}^{1}D$ through the fourth, is
\begin{equation}
\mathcal{M}^{(2)}
=
\begin{pmatrix}
I & {}^{1}D & ({}^{1}D)^2\\
{}^{1}D & ({}^{1}D)^2 & ({}^{1}D)^3\\
({}^{1}D)^2 & ({}^{1}D)^3 & ({}^{1}D)^4
\end{pmatrix}
\succeq 0.
\end{equation}
To obtain a semidefinite-programming relaxation~\cite{Vandenberghe_SDP_1996,Mazziotti_SDP,Mazziotti_SDP_2}, the powers of the 1-RDM are replaced by independent matrix variables $X_k$, with $X_0=I$ and $X_1={}^1D$, giving
\begin{equation}
\widetilde{\mathcal{M}}^{(2)}
=
\begin{pmatrix}
X_0 & X_1 & X_2\\
X_1 & X_2 & X_3\\
X_2 & X_3 & X_4
\end{pmatrix}
\succeq 0.
\end{equation}
This corresponds to a truncated Hausdorff moment formulation on the occupation interval [0,1]~\cite{ShohatTamarkin1943,Lasserre2010}.  In terms of these lifted moments, the second-order correction becomes
\begin{equation}
C[{}^{1}D]
=
-w_0\,\mathrm{Tr}\!\left(X_1-X_2\right)
+
u_0\,\mathrm{Tr}\!\left(X_2-2X_3+X_4\right).
\end{equation}
Thus, the relaxation transforms the quartic polynomial correction into a linear functional of the lifted density-matrix moments.

Consistency of the lifted moments with the physical occupation interval is further imposed through the particle-hole localizing matrix,
\begin{equation}
\widetilde{\mathcal{L}}_{DQ}^{(1)}
=
\begin{pmatrix}
X_1-X_2 & X_2-X_3\\
X_2-X_3 & X_3-X_4
\end{pmatrix}
\succeq 0.
\end{equation}
Together with the positive-semidefinite moment matrix, this localizing condition constrains the lifted variables to be \textcolor{black}{moments of a distribution of occupations on $[0,1]$}, and hence, for the convex particle-hole functional considered here, these conditions recover the physical moments $X_k=({}^{1}D)^k$ at the optimum. Details of the moment relaxation and its exactness are provided in the Supplemental Material.

As in Kohn--Sham density functional theory, the 1-RDM functional is
solved self-consistently. At each iteration, the conventional
Kohn--Sham Fock matrix, ${}^{1}F_{\mathrm{KS}}$, is constructed from the
current density, and the 1-RDM is obtained from the semidefinite program
\begin{equation}
\min_{{}^{1}D\textcolor{black}{,\,X_2,\,X_3,\,X_4}}\,
\left\{
\mathrm{Tr}\!\left[
{}^{1}F_{\mathrm{KS}}\,{}^{1}D
\right]
+
C[{}^{1}D]
\right\}
\end{equation}
subject to
\begin{equation}
{}^{1}D \succeq 0,\qquad
{}^{1}Q \succeq 0,\qquad
\widetilde{\mathcal{M}}^{(2)} \succeq 0,\qquad
\widetilde{\mathcal{L}}_{DQ}^{(1)} \succeq 0,
\end{equation}
and $\mathrm{Tr}({}^{1}D)=N$. The optimized 1-RDM generates an updated
density from which a new Kohn--Sham Fock matrix is constructed. The
Kohn--Sham and semidefinite-programming steps are repeated to
self-consistency. Because the optimization involves only one-particle
matrices, the computational scaling remains comparable to that of
conventional DFT, scaling as $\mathcal{O}(r^3)$ with the size $r$
of the one-particle basis.

\color{black}
Because the correction depends only on the natural-orbital occupations, the optimal 1-RDM commutes with ${}^{1}F_{\mathrm{KS}}$, and the semidefinite program yields Kohn--Sham-orbital occupations satisfying $\varepsilon_i+c'(n_i)=\mu$, where $\varepsilon_i$ are the orbital energies and $\mu$ enforces $\mathrm{Tr}({}^{1}D)=N$. For convex corrections each occupation solves a monotone scalar equation, which at second order is a cubic whose unique real root follows from Cardano's formula (SM).  For $|\varepsilon_i-\mu|<w_0$ the occupations are $n_i=\frac12[1-(\varepsilon_i-\mu)/w_0]$ for $u_0=0$ and $n_i=\frac12\{1-[(\varepsilon_i-\mu)/w_0]^{1/3}\}$ for $u_0=2w_0$, and outside this window they are 0 or 1.  The semidefinite solutions agree with these closed-form occupations to within $2\times10^{-7}$ for N$_2$.  The thermal (entropic) correction $\theta\sum_i[n_i\ln n_i+(1-n_i)\ln(1-n_i)]$ of TAO-DFT and i-DMFT~\cite{Chai_DFT_Fractional_2012,JWang_aps_2022}, which yields Fermi--Dirac occupations, is a special infinite-order case of Eq.~(\ref{eq:am}), with coefficients $a_{2j}=-4\theta\sum_{k>j}[2k(2k-1)]^{-1}$ that decay as $1/j$ (SM).
\color{black}

\noindent {\em Results---} We apply ph-RDMFT as a correction to SCAN-DFT~\cite{Sun_SCAN_2015}
in the correlation-consistent polarized valence double-zeta (cc-pVDZ)~\cite{Dunning1989} basis set. For the calculations presented below, we parameterize the first- and second-order coefficients introduced above as
$w_0=\kappa\gamma$ and $u_0=f\kappa\gamma$, respectively, where $\kappa$
is the damping parameter used in our previous RDMFT
formulation~\cite{Gibney_JPCL_2022,Gibney_jctc2025_200xc}, $\gamma$ is a
system-dependent interaction parameter determined from the trace of the
two-electron repulsion matrix with a size-extensive
renormalization~\cite{Gibney_PRL_2023,Gibney_PRA_2024}, and $f$ controls
the magnitude of the second-order particle-hole correction. We employ the
previously established default value \textcolor{black}{$\kappa=0.15$} and the maximum globally
convex value, $f=2$. PySCF~\cite{Sun_PySCF_2018} is used as the DFT solver, while the convex
RDMFT optimization is implemented using the SCS~\cite{ODonoghue_SCS_2016} solver in CVXPY~\cite{Diamond_CVXPY_2016}\textcolor{black}{ with a convergence tolerance of $10^{-10}$.
Variational 2-RDM calculations (v2RDM)~\cite{Mazziotti_SDP, Fukuda.2007, Gidofalvi_Mazziotti_2008, Mazziotti_SDP_2, Mazziotti_V2RDM_2016, DePrince_Geoemtry_2019_jctc, Knight.2022, Mazziotti_Nrep_2023, Gao.2025dgs, Schouten.20250j, Mazziotti.2026} for complete active spaces are performed with the Quantum Chemistry Toolbox for Maple~\cite{Montgomery2020QCP, QCT2026, Maple2026}.  We report natural-orbital occupations in the figures for spatial orbitals, which range from 0 to 2}. We
apply the RDMFT-corrected SCAN functional to two classes of systems:
periacenes of increasing size and Fe-porphyrin. In each case, we compare
the natural-orbital occupation numbers obtained from the second-order
particle-hole expansion ($f=2$) with those from the original RDMFT
formulation ($f=0$) and with multireference reference data.

\begin{figure}[t]
   \centering
   \includegraphics[width=0.95\columnwidth]{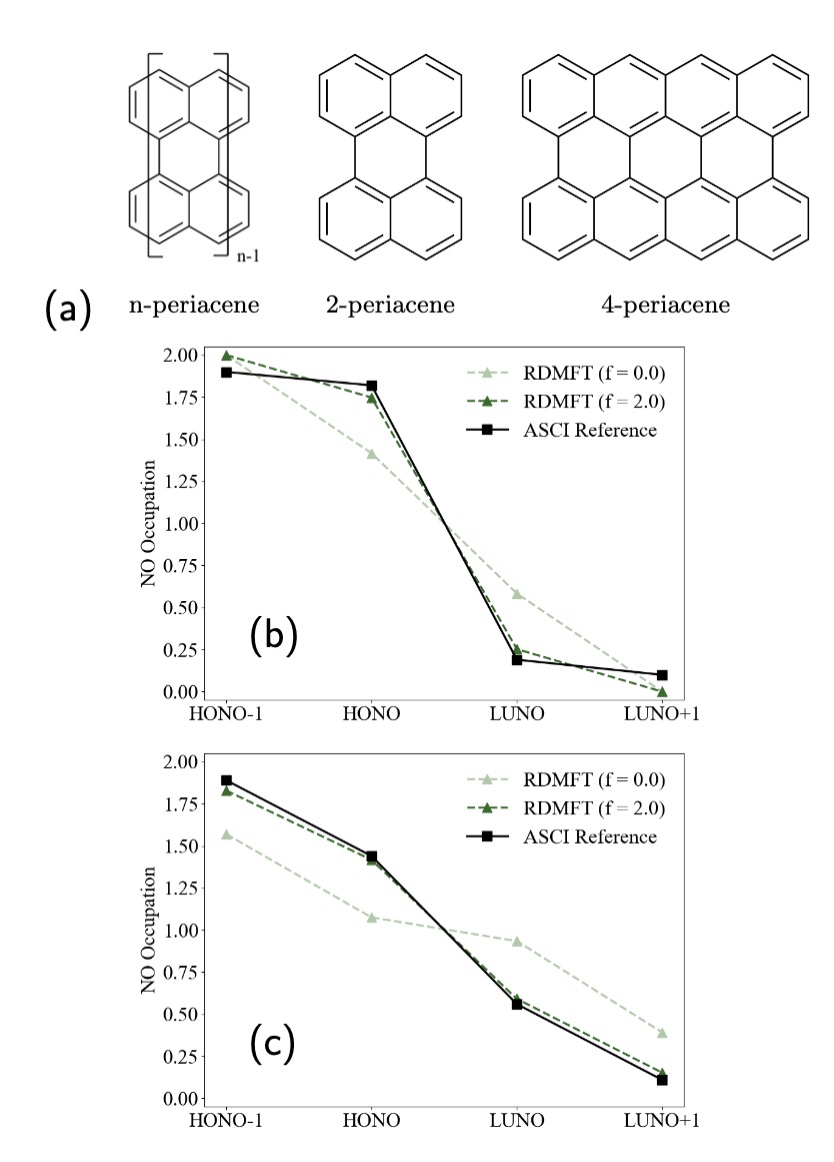}
   \caption{(a) Molecular structure of n-periacene, 2-periacene, and 4-periacene.\textcolor{black}{ Natural-orbital occupations of (b) 2-periacene and (c) 4-periacene} computed at the RDMFT-SCAN/cc-pVDZ level of theory.  $f=0$ denotes the original RDMFT, and $f=2$ denotes RDMFT including higher-order corrections. The RDMFT calculations are performed with $\kappa = 0.15$. The ASCI data \textcolor{black}{are} taken from Ref.~\onlinecite{HG_ASCI}.}
    \label{fig:rdmft_scan_ccpvdz}
\end{figure}

We first consider a series of periacenes of increasing size. Periacenes
are two-dimensional analogues of acenes, consisting of two parallel acene
chains fused together to form an extended $\pi$-conjugated framework.
Their electronic structure has attracted considerable
interest~\cite{DePrince_Geoemtry_2019_jctc,plasser_angew_periacene_2013,
scuseria_periacene_2013,HG_ASCI,periacene_Yanai_2013_jctc}
because they serve as finite models of graphene and exhibit increasing
polyradical character with molecular size, making them valuable benchmark
systems for methods designed to describe strong electron correlation. As
reference data, we use the adaptive sampling configuration interaction
(ASCI) natural-orbital occupations reported by Head-Gordon and
co-workers~\cite{HG_ASCI}.
The periacene geometries are taken
from Ref.~\onlinecite{DePrince_Geoemtry_2019_jctc}. For the systems
examined, the original RDMFT overestimates the degree of orbital
fractionalization. In contrast, the particle-hole moment expansion yields
occupation numbers in excellent agreement with the ASCI reference for
\textcolor{black}{2- and 4-periacene (Fig.~\ref{fig:rdmft_scan_ccpvdz})}. Results for periacenes with an odd number of repeating
units are provided in the SM. We additionally verify
in the SM that the total energy scales linearly with
the increasing number of periacene units.

\begin{figure}[t]
   \centering
   \includegraphics[width=\columnwidth]{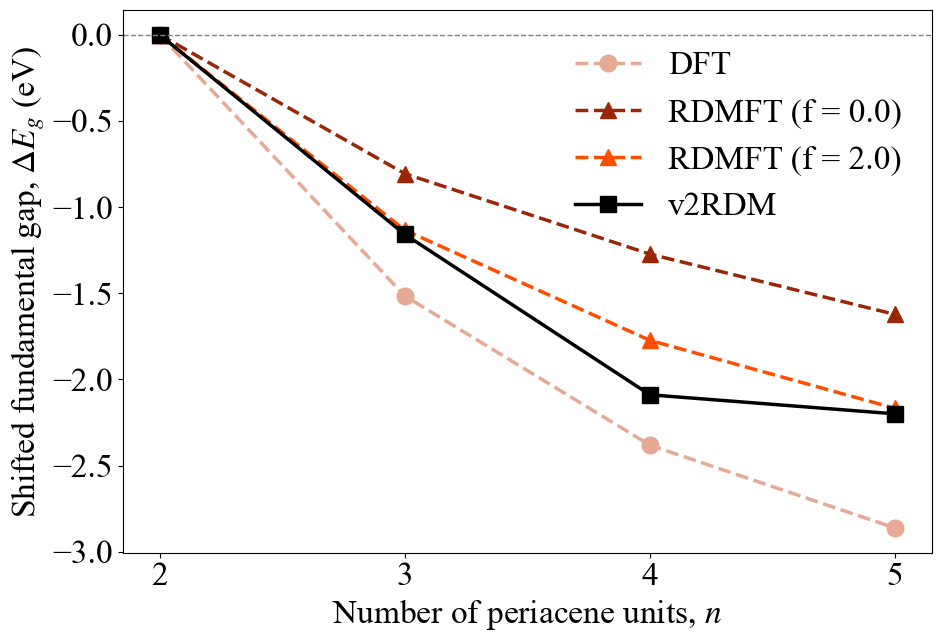}
   \caption{Fundamental \textcolor{black}{gap} of n-periacene computed at the RDMFT-SCAN/cc-pVDZ level of theory compared against fundamental \textcolor{black}{gap} calculated with v2RDM. The RDMFT calculations are performed with $\kappa = 0.15$. The $\pi$ molecular orbitals are chosen as active orbitals in the v2RDM calculations.}
    \label{fig:bandgap}
\end{figure}

Together with the natural occupation numbers, we compute the fundamental \textcolor{black}{gaps}~\cite{Sharma_bandgap_prb} for the periacene series, \textcolor{black}{$E_g=E_{N+1}+E_{N-1}-2E_N$,} where $E_{N+1}$, $E_{N-1}$ and $E_{N}$ denote the energies of the anion, cation\textcolor{black}{,} and \textcolor{black}{neutral species, respectively}. \textcolor{black}{The neutral periacenes are treated as spin singlets, while the anion and cation are treated as spin doublets with $\mathrm{Tr}({}^{1}D)=N\pm1$. The parameter $\gamma$, which sets both coefficients, is recalculated for each charge state.} As we see from Fig.~\ref{fig:bandgap}, the fundamental \textcolor{black}{gap} computed with RDMFT ($f=2$) is in better agreement with the v2RDM reference, using a complete active space of the $\pi$ molecular orbitals, than those obtained with DFT or RDMFT with $f=0$\textcolor{black}{, with mean absolute deviations of 0.09, 0.33, and 0.44~eV for RDMFT ($f=2$), SCAN, and RDMFT ($f=0$), respectively.}  This improved agreement suggests that accurately describing the fractional orbital occupations, and hence the increasing multireference character across the periacene series, is important for predicting fundamental \textcolor{black}{gaps}.

\begin{figure}[t]
   \centering
   \includegraphics[width=0.95\columnwidth]{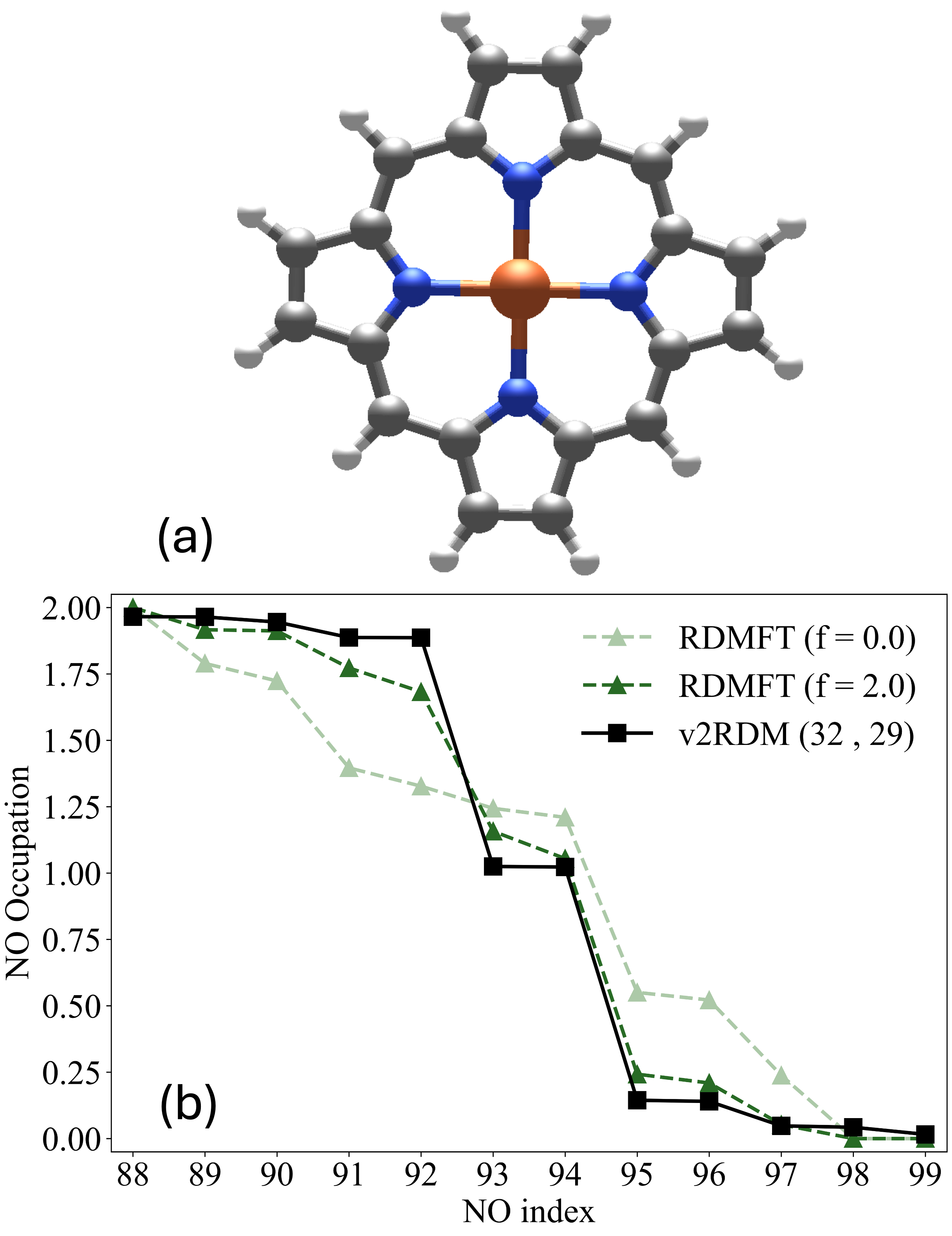}
   \caption{(a) Molecular structure and (b) natural-orbital occupations of Fe-porphyrin computed at the RDMFT-SCAN/cc-pVDZ level of theory. $f=0$ denotes the original RDMFT, and $f=2$ denotes RDMFT including the higher-order correction. The RDMFT calculations are performed with $\kappa = 0.15$.}
   \label{fig:Fe_P_rdmft_v2rdm}
\end{figure}

As our final system, we consider the natural-orbital occupations of
Fe-porphyrin. As the central component of the heme cofactors in proteins
such as hemoglobin, myoglobin, and catalase, Fe(II) porphyrin (Fe(P)) has
long served as a challenging benchmark for correlated electronic-structure
methods.  Its low-lying spin states have been investigated using
high-level correlated and natural-orbital functional methods,
including HCISCF~\cite{sharma_porphyrin_2017},
DMRG~\cite{chan_porphyrin_2015}, ASCI~\cite{HG_ASCI}, and
natural-orbital functional theory~\cite{Piris_FeP_2023}. We perform RDMFT calculations using the optimized triplet geometry from
Ref.~\onlinecite{Groenhof_porphyrin_2005} with the SCAN functional and
the cc-pVDZ basis set. \textcolor{black}{The triplet state is treated using an unrestricted Kohn–Sham (UKS) reference, allowing distinct $\alpha$- and $\beta$-spin orbitals and one-particle reduced density matrices (1-RDMs), which are subsequently used to initialize the RDMFT calculations.} Figure~\ref{fig:Fe_P_rdmft_v2rdm} compares the frontier
natural-orbital occupations obtained with the original RDMFT functional
($f=0$) and the second-order particle-hole correction ($f=2$) against a
v2RDM calculation using a (32,29) active space. The original RDMFT
functional produces substantially greater fractionalization than the
v2RDM reference, with pronounced underoccupation of the frontier occupied
orbitals (NO indices 89--92) and overoccupation of the low-lying virtual
orbitals (NO indices 95--97). Inclusion of the second-order correction
markedly reduces these deviations: the occupied-orbital occupations shift
toward 2, while the virtual-orbital occupations shift toward 0, bringing
the RDMFT results substantially closer to the v2RDM reference. In contrast,
the occupations of the two approximately singly occupied orbitals (NO
indices 93 and 94) are affected much less by the correction. Results for
values of $f$ beyond the globally convex bound of $f=2$ are provided in
the \textcolor{black}{SM}, where \textcolor{black}{the semidefinite relaxation evaluates the convex envelope of the correction and} further increasing $f$ leads to even
closer agreement with the v2RDM reference.
These results demonstrate that the second-order particle-hole term suppresses the
excessive orbital fractionalization of the leading-order functional while
retaining the strongly correlated character of the frontier orbitals.

As an additional benchmark, we examine N$_2$ bond dissociation~\cite{Pernal_N2_2019}. Comparable agreement with complete active-space self-consistent-field (CASSCF) calculations is obtained with the leading-order functional at smaller $\kappa$ and with the second-order functional at larger $\kappa$, demonstrating the additional flexibility introduced by the higher-order particle-hole term (SM).

\noindent {\em Conclusions---}In this work, we introduced a particle-hole expansion that systematically
generalizes the 1-RDM correction to density functional approximations.
The previously derived cumulant-based correction emerges as the leading
term of the expansion, while higher-order particle-hole moments provide
additional flexibility in the dependence of the functional on the
natural-orbital occupations. For the second-order expansion, we derived
bounds on the expansion coefficients from convexity and formulated the
resulting optimization in terms of density-matrix moments that can be
solved by semidefinite programming. Within these bounds, the optimization
yields a closed-form expression for the natural-orbital occupations.
Applications to conjugated polyradical systems and Fe-porphyrin demonstrate
that the additional particle-hole term can substantially improve the
natural-orbital occupations when the leading-order functional produces
excessive fractionalization.

The particle-hole expansion occupies a distinct position between
conventional density functional approximations and traditional RDMFT~\cite{Su_DFT_RDMFT_2021}.
Unlike conventional Kohn--Sham DFA~\cite{KS_DFT,WYang_SC_2008,
Hollett_Gill_2011}, it introduces an explicit
correlation correction that depends on a generally non-idempotent 1-RDM
and can therefore describe strongly fractional natural-orbital
occupations. Unlike conventional RDMFT~\cite{Muller_1984,Goedecker_Umrigar_1998,Piris_2006,Piris_NOF_2013,Piris_Global_2017,Piris_Ugalde_2014,Piris_GNOF_2021,Piris_Perspective_2024,
JWang_aps_2022,Schilling_PRL_2019,Liebert_FunctionalTheory_2023,Schilling_iDMFT_2026,Matz_thermal_2026,JWang_21}, however, the particle-hole
functional is constructed as a correction to an existing DFA, allowing
the accuracy and efficiency of established density functionals to be
retained. Related strategies combining
density-functional and density-matrix-functional descriptions have also
been explored previously~\cite{Rohr_DFT_RDMFT_2010,Gibney_JCTC_2022}.
\textcolor{black}{Recent thermal (entropic) approaches such as TAO-DFT and i-DMFT~\cite{Chai_DFT_Fractional_2012,JWang_aps_2022,Gibney_JCTC_2022,Ramos_Pavanello_2019,Martinez_Pavanello_2023,Matz_thermal_2026,Han_FTACRDMFT_2026,Graf_temperature_2026} fall within the particle-hole expansion as an infinite-order case whose coefficients are fixed by a single temperature-like parameter, whereas the expansion, by treating the coefficients independently, can adjust the shape of the occupation dependence, as the second-order term illustrates. Unlike the entropic correction, whose diverging derivative at $n=0,1$ mirrors the exchange force of the exact functional~\cite{Schilling_PRL_2019}, finite truncations yield integer occupations away from the Fermi level, confining the correction to static correlation and avoiding double counting of the dynamic correlation already described by the DFA.}

Although the method requires a semidefinite optimization, in
the natural-orbital basis\textcolor{black}{, which coincides with the Kohn--Sham basis,} the moment and localizing conditions reduce to
small $3\times3$ and $2\times2$ positive-semidefinite blocks for each
occupation.  This block structure permits an $\mathcal{O}(r^3)$ implementation,
comparable in scaling to conventional DFT\textcolor{black}{,
and yields a closed-form expression for the occupations at second order}. The resulting framework provides a
systematic, size-extensive hierarchy of increasingly expressive density-matrix
corrections for strong electron correlation within density functional theory.

\begin{acknowledgments}
D.A.M. gratefully acknowledges the U.S. National Science Foundation Grant No. CHE-2155082 for support.  AI tools Claude Opus 5.5 High and ChatGPT 5.6 High were used for scientific discussions and manuscript editing. All scientific content was independently evaluated and verified by the authors, who take full responsibility for the results and conclusions.
\end{acknowledgments}

\bibliography{allrefs,new_refs}

\foreach \x in {1,...,18}{%
  \clearpage
  \includepdf[pages={\x}]{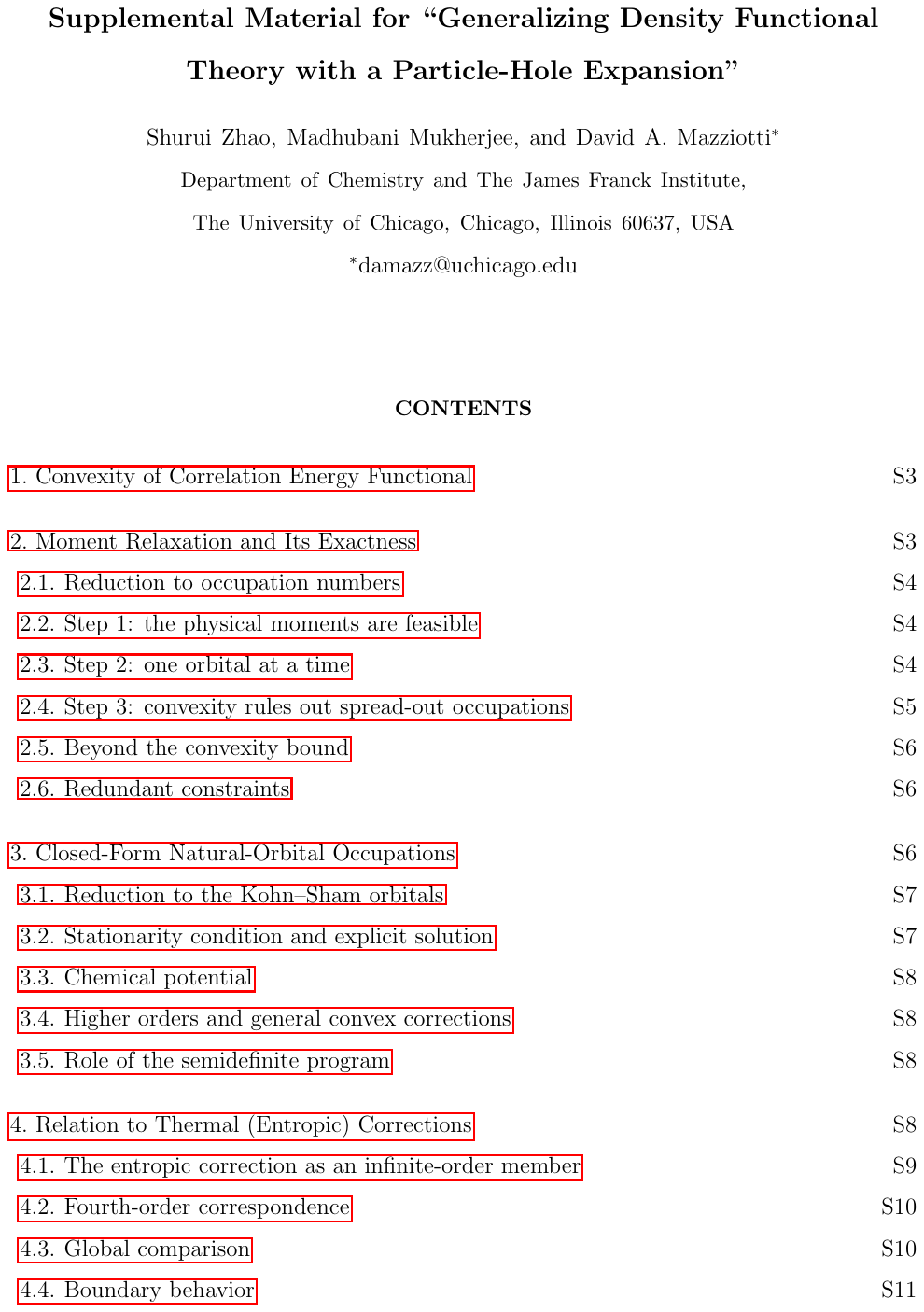}
}

\end{document}